\documentclass[envcountsame]{llncs}

\usepackage{amssymb}
\usepackage{graphicx}
\usepackage{booktabs}
\usepackage{pdflscape}
\usepackage{mdframed}
\usepackage{subfig}
\usepackage{mathtools}
\usepackage{lineno}
\let\doendproof\endproof
\renewcommand\endproof{~\hfill\qed\doendproof}

\newcommand{\NP}{\textsc{NP}}

\newcommand{\problembox}[4]{
	\begin{framed}
		{\sc #1} \\
		\begin{tabular}{p{.18\textwidth} p{.82\textwidth}}
			\hfill \bf Input: & {#2}\\
			\hfill \bf Question: & {#4}\\
		\end{tabular}
	\end{framed}
}
\newcommand{\ec}{\textsc{Equitable Coloring}}
\newcommand{\lck}{\textsc{List $k$-Coloring}}
\newcommand{\pe}{\textsc{Precoloring Extension}}
\newcommand{\gc}{\textsc {Graph Coloring}}
\newcommand{\fpt}{\textsc {FPT}}
\usepackage{xspace}
\usepackage{framed}

\begin{document}
\title{Parameterized Coloring Problems on Threshold Graphs }
%
\author{I. Vinod Reddy}
\authorrunning{I. Vinod Reddy}
%
\institute{Indian Institute of Technology Bhilai, India}

\maketitle

\begin{abstract}
In this paper, we study several coloring problems on graphs from the viewpoint of parameterized
complexity. We show that \pe{} is fixed-parameter tractable ($\fpt$) parameterized by distance to clique. We show that \ec{} is $\fpt$ parameterized by the distance to threshold graphs. We also study the \lck{} and show that the problem is NP-complete on split graphs and it is $\fpt$ parameterized by solution size on split graphs. 
\end{abstract}

\section{Introduction}

Given a graph $G$, and a positive integer $k$, the $k$-coloring problem is to color the vertices of $G$ with at most $k$ colors such that adjacent vertices receive different colors. 
This is a well studied problem in computer science due to its theoretical and practical applications. The problem is $\NP$-complete for every fixed $k \geq 3$. The problem is well studied from the perspective of parameterized complexity. For example, it is  $\fpt{}$ when parameterized by vertex cover \cite{bodlaender2014kernelization}, 
tree-width~\cite{courcelle1990monadic} and distance to clique~\cite{saether2016between}. On the other hand, it is W[1]-hard when parameterized by clique-width~\cite{fomin2010intractability} and  distance to split graphs~\cite{cai2003parameterized}. 

In this paper we study the parameterized complexity of several graph coloring problems (\pe{}, \ec{}, \lck{}), with respect to distance parameters~\cite{bulian2017parameterized,guo2004structural}, where we take the parameter to be the distance to threshold graphs. It is an intermediate parameter between vertex cover and clique-width. Many variants of graph coloring are fixed-parameter tractable when parameterized by the vertex cover. However, the parameter vertex cover is very restrictive in the sense that the class of graphs with bounded vertex cover is small. The parameter distance to threshold graphs generalizes vertex cover in the sense that the class of graphs with bounded vertex cover contains in the class of graphs with bounded distance to threshold graphs. Thus the existence of $\fpt{}$ algorithms for any problem parameterized by
distance to threshold graphs supplants the results obtained by parameterizing with vertex cover. Hence this reduces the gap between tractability and intractability.

\paragraph{Problems considered.}
The problems we consider in this paper are as follows.

\problembox{\pe{}}{A graph $G$, an integer $r$, a subset $W \subseteq V(G)$ and a  precoloring $C_W: W \rightarrow [r]$}{The size $k:= |X|$ of the modulator to $\mathcal{F}$.}{ Is there a proper coloring $c: V(G) \rightarrow [r]$ of $G$ such that $c(u)=C_W(u)$ for every $u \in W$?}

\problembox{\ec{}}{A graph $G$ and an integer $r$ }{The size $k:= |X|$ of the modulator to $\mathcal{F}$.}{Is there a proper  coloring of $G$ using at most $r$ colors such that the sizes of any two color classes differ by at most one?}


\problembox{\lck{}}{A graph $G$ and an assignment $L: V(G) \rightarrow S \subseteq[k]$ of color lists to the vertices of $G$.
	}{The size $k:= |X|$ of the modulator to $\mathcal{F}$.}{Is there a proper coloring $c: V(G) \rightarrow [k]$ such that $c(u) \in L(u)$ for every $u \in V(G)$?}

\paragraph{Related work.}
Both \pe{} and \ec{} are $\fpt$ when parameterized by the vertex cover number~\cite{fiala2011parameterized} and $W[1]$-hard when parameterized by tree-width~\cite{fellows2011complexity}. However they can be solved in polynomial time on graphs of bounded tree-width~\cite{bodlaender2005equitable,jansen1997generalized}. Ganian~\cite{ganian2015improving} showed that both problems are $\fpt$ parameterized twincover. Doucha et al.~\cite{doucha2012cluster} showed that \pe{} is $\fpt$ parameterized by bounded cluster vertex deletion and $W[1]$-hard parameterized by unbounded cluster vertex deletion. They also showed that \ec{} is $\fpt$ parameterized by unbounded cluster vertex deletion. D.Marx~\cite{marx2006parameterized} showed that \pe{} is $W[1]$-hard parameterized by either distance to interval graphs or distance to chordal graphs. 
\lck{} is $W[1]$-hard when parameterized by the vertex cover number. The problem remains hard even for split graphs. Banik et al.~\cite{banik2019fixed} showed that $(n-k)$- regular list coloring is $\fpt$ parameterized by $k$. Jansen et al.~\cite{jansen2013data} showed that $q$-regular list coloring is $\fpt$ parameterized by combined parameter $q+k$.

\paragraph{Our contributions.}
We summarize our results below.
\begin{itemize}
	\item In Section~3, we show that \pe{} is $\fpt$ parameterized by distance to clique.
	\item In Section~4, we show that \ec{} is  $\fpt$ parameterized by distance threshold graphs.
	\item In Section~5, we  show that \lck{} is (a) NP-complete on split graphs and (b) $\fpt$ parameterized by $k$ on split graphs.
\end{itemize}
\section{Preliminaries}

In this section, we introduce some basic notation and terminology related to graph theory and parameterized complexity. For $n \in \mathbb{N}$, we use $[n]$ to denote the set $\{1,2, \dots, n \}$.

\paragraph{Graph theory.} All graphs we consider in this paper are undirected, connected, finite, and simple. For a  graph $G=(V,E)$, by $V(G)$ and $E(G)$ we denote the vertex set and edge set of $G$ respectively. We use $n$ to denote the number of vertices and $m$ to denote the number of edges of a graph.
An edge between two vertices $x$ and $y$ is denoted as $xy$ for simplicity. 
For a  subset $X \subseteq V(G)$, the graph $G[X]$ denotes the subgraph of $G$ induced by vertices of $X$. Also, for simplicity, we use $G \setminus X$ to refer to the graph obtained from $G$ after removing the vertex set $X$. 
For a vertex $v\in V(G)$,
by $N(v)$ we denote the set $\{u \in V(G) ~|~ vu \in E(G)\}$ and we use $N[v]$ to denote the set $N(v) \cup \{v\}$. The neighborhood of a vertex subset $S \subseteq V(G)$ is $N(S)= (\cup_{v \in S} N(v)) \setminus S$. A vertex is called \emph{universal vertex} if it is adjacent to every other vertex of the graph.  A graph $G$ has deletion distance $d$ to a graph class $\mathcal{F}$ if there exists a set $X \subseteq V(G)$ of $d$ vertices such that $G \setminus X \in \mathcal{F}$. We say that $X$ is an $\mathcal{F}$-modulator of graph $G$. 

A graph is a \emph{split graph} if its vertices can be partitioned into a clique and
an independent set.
A graph is a \emph{threshold graph} if it can be constructed from the one-vertex graph by repeatedly
adding either an isolated vertex or a universal vertex. 
The class of threshold graphs is the intersection of split graphs and cographs~\cite{mahadev1995threshold}.
We denote a threshold graph as $G = (C, I)$, where $(C,I)$ denotes the partition of $G$ into a clique and an independent set, respectively.
It is easy to see that every induced subgraph of a threshold graph is also a threshold graph. We have the following characterization of threshold graphs: A graph $G$ is a threshold graph if and only if it is $(P_4, C_4, 2K_2)$-free. For any two vertices $x, y$ in a threshold graph $G$ we have either $N(x) \subseteq N[y]$ or $N(y) \subseteq N[x]$ (neighborhood containment property). For more details on standard graph-theoretic notation and terminology, we refer the reader to~\cite{diestel2005graph}.

As threshold graphs are $(P_4, C_4, 2K_2)$-free,  checking whether a given graph $G$ has vertex deletion distance $d$ to the class of threshold graphs is fixed-parameter tractable. Therefore without loss of generality, in this paper, we assume that threshold graph modulator is given as a part of the input. 

\begin{figure}
	\centering
	\includegraphics[scale=0.9,trim={0cm 13.4cm 9.3cm 12.3cm},clip]{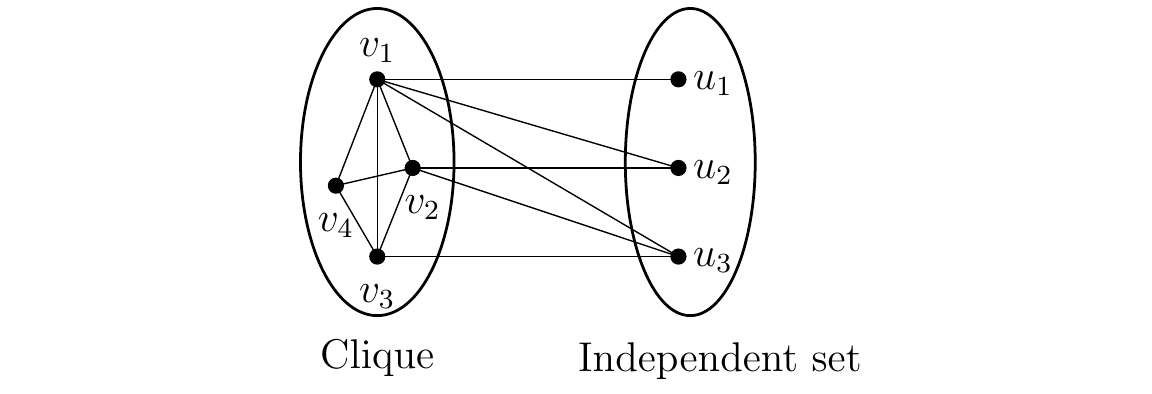}
	\caption{An example of a threshold graph $G=(C,I)$, where $C=\{v_1,v_2,v_3,v_4\}$ and $I=\{u_1,u_2,u_3\}$.  The vertex $v_1$ is universal in $G$. We can also see that $N[v_4] \subseteq N[v_3] \subseteq N[v_2]\subseteq N[v_1]$  and $N(u_1) \subseteq N(u_2) \subseteq N(u_3)$.}
	\label{fig-threshold}
\end{figure}

\paragraph{Parameterized complexity.} A parameterized problem denoted as $(I,k)\subseteq \Sigma^*\times \mathbb{N}$, where $\Sigma$ is fixed alphabet and $k$ is called the parameter. We say that the problem $(I,k)$ is {\it fixed parameter tractable} with respect to parameter $k$ if there exists an algorithm which solves the problem in time $f(k) |I|^{O(1)}$, where $f$ is a computable function. 
A \emph{kernel} for a parameterized problem $\Pi$ is an algorithm which transforms an instance $(I,k)$ of $\Pi$ to an equivalent instance $(I',k')$ in polynomial time such that $k' \leq k$ and $|I'| \leq f(k)$  for some computable function $f$. 
For more details on parameterized complexity, we refer the reader to the texts~\cite{CyganFKLMPPS15,downey2013fundamentals}.

\section{Precoloring Extension}

In this section,  we show that \pe{} is $\fpt$ when parameterized by the distance to clique.

%

\begin{theorem} \label{Th-peclqiue}
	\pe{} can be solved in $O(2^{k (3\log k +3)}r(n+\sqrt{nm}))$ time when parameterized by the  distance  to clique.
\end{theorem}
\begin{proof}
	Let $X \subseteq V(G)$ of size $k$ such that $G \setminus X=C$ is a clique. Let $f: W \subseteq V(G) \rightarrow [t]$ be the given precoloring, where $t \leq r$ and $f(W)=[t]$. 
	Let $V_C= W \cap C$ and $V_X= W \cap X$ be the set of precolored vertices in the clique $C$ and the modulator $X$ respectively. 
	Let $S_C=f(V_C)$ and $S_X=f(V_X)$.
	As $C$ is a clique, no color of $S_C$ can be used to color any uncolored vertices of the clique.  
	
	First, we check whether it is possible to use some colors of $S_C \cup S_X$ to color vertices of $X \setminus V_X$.
	For each vertex in $X \setminus V_X$ we assign a list of colors according to their neighborhood in $G$, i.e., for each $u \in X \setminus V_X$, $L(u):= [t] \setminus  f(N(u) \cap W)$.
	Let $X_h=\{v \in X \setminus V_X ~|~ |L(v)|>2k\}$.
	Any vertex $v$ in $X_h$ can be colored at the end :  
	$L(v)$ has at least $k$ colors from $S_C$, and there are at most $k$ uncolored vertices in $X$. Therefore $v$ can be colored greedily with one of the color from $L(v)$ at the end.
	
	We are left with a subset $Y\subseteq X \setminus V_X$ of at most $k$ vertices having at most $2k$ colors in their lists.
	Partition $Y=Y' \cup Y''$ into two sets such that $Y'$ contains vertices which gets the colors from $[t]$
	and $Y''$ contains vertices which gets colors from the set $[r] \setminus [t]$.  Since $|Y'|\leq k$ and $|L(v)|\leq 2k$ for all $v \in Y'$,
	we try all possible $O(k^{2k})$ ways to color the vertices of $Y'$. Similarly as  the size of $Y''$ is at most $k$, we can assign colors to the vertices of $Y''$ from the set $[r] \setminus [t]$ in $O(k^k)$ time. 
	
	Now we have a partial coloring of $G$ in which all vertices of $X \setminus X_h$ and some vertices of $C$ are colored. 
	The uncolored vertices in the clique are colored by finding a maximum matching in the following bipartite graph. 
	The vertex set of the bipartite graph contains all $r$ colors as one partition and uncolored clique vertices as the other. A color $c$ is adjacent to a vertex $x$, if $x$ is not adjacent to a vertex colored with $c$ in $G$. 
	If there is a matching saturating all uncolored vertices of $C$, then we assign the colors to clique as per the matching.  In the end, we greedily assign colors to the vertices of $X_h$ from their lists.  
	
	\paragraph{Running time.} Computing the threshod graph modulator $X$ takes $O(4^k(m+n))$ time. Trying all possible ways of partitioning $Y$ into $Y'$ and $Y''$  takes $O(2^k)$ time and coloring the vertices of $Y$ requires $O(k^{3k})$ time. Constructing the bipartite graph takes $O(r(n+m))$ time. 
	Computing the maximum matching in a bipartite graph need $O(n+\sqrt{nm})$ time. 
	Therefore running time of the whole algorithm is $O(2^{k (3\log k +3)}r(n+\sqrt{nm}))$.
\end{proof}

\section{Equitable Coloring}
In this section, we first show that \ec{} is solvable in polynomial time for the
class of threshold graphs. Next, we describe $\fpt$ algorithm for the problem when parameterized by distance to threshold graphs.

\subsection{Threshold graphs}
Chen et al.~\cite{chen1995equitable} showed that \ec{} can be solved in polynomial time on split graphs. However, here we present a simple polynomial-time algorithm for \ec{} on threshold graphs. This will give a useful warmup for the next part, where we describe our FPT algorithm for the
problem. 
	\begin{lemma}
		\ec{} can be solved in polynomial time on threshold graphs.
	\end{lemma}
		\begin{proof}
			Let $G=(C,I)$ be a threshold graph. As $G$ has at least one universal vertex, the size of a color class in any equitable coloring of $G$ is at most two. If $r <|C|$ then the given instance is a \textsc{No} instance as we need at least $|C|$ colors to color the clique. Therefore without loss of generality, we assume that $r \geq |C|$. 
			For each integer $t$ with $|C| \leq t \leq r$, we test whether $G$ has an equitable coloring with $t$ colors.  
			Given an integer $t$, first color the vertices of the clique $C$ using $|C|$ many colors.  Now we order the vertices of the independent set according to their degree from highest degree to lowest degree. Assign the $t-|C|$ unused colors to the first $t-|C|$ independent vertices according to the above ordering (this greedy choice works because of neighborhood containment property of the threshold graphs). So far, we have used each of the $t$ colors exactly once. Since the size of the maximum color class in any equitable coloring is at most two, any color can be used at most once to color the rest of the uncolored independent set vertices. We do this by solving a network flow probem as follows. 
			
			We add a source vertex  and $t$ vertices representing colors and  connect source vertex with $t$ vertices with each having the capacity one. Next, we create a vertex to represent every uncolored independent set vertex of $G$, and add edges from these vertices to the sink with capacity one. Finally, we add edges of capacity one from $t$ color vertices to independent set vertices if the color may be assigned to that vertex. Then compute the maximum flow in the above-constructed graph and check whether it is equal to the number
			of uncolored vertices in $I$ and if this is the case,
			we immediately obtain a solution in $G$. Since the maximum flow is bounded by the number of vertices, this flow problem
			can be solved in time $O(mn)$. Altogether we get a polynomial-time algorithm for the problem.
		\end{proof}
\subsection{Parameterized by distance to threshold graphs}

 The $\fpt$ algorithm is very involved. We give a brief overview of the main ideas in our algorithm. Let $G$ be a graph and $X \subseteq V(G)$ of size at most $k$ such that $G \setminus X=(C,I)$ is a threshold graph. We start by guessing the coloring of $X$ in a solution and then  try to extend it to an equitable coloring of $G$. To extend a coloring of $X$ to $G$ we use the following key ideas (a) In any equitable coloring of $G$ the size of any color class is at most $k+2$. (b) As the size of $X$ is at most $k$, we can guess the color class sizes of colors used to color vertices of $X$. (c)  We use the neighborhood containment property of threshold graphs to assign new colors (colors not used in $X$) to color the clique and the independent set of $G \setminus X$ respectively.

	\begin{theorem}
		\ec{} is fixed-parameter tractable when parameterized by distance to threshold graphs.
	\end{theorem}

	\begin{proof} 
		Let $X \subseteq V(G)$ of size $k$ such that $G \setminus X=(C,I)$ is a threshold graph. As $G \setminus X$ has at least one universal vertex $u$, in any equitable coloring of $G$, the color of $u$ is unique in $G \setminus X$. Therefore the color of $u$ can appear at most $k+1$ times in $G$. So the maximum size of a color class is at most $k+2$ in an equitable coloring of $G$. An $r$-equitable colorable graph may not be $(r+1)$-equitable colorable, therefore for all possible values $t \in [r]$ we check whether $G$ is $t$-equitable colorable. 
		If the number of colors used in an equitable coloring of $G$  is $t$ then we can find the number of color classes of
		size $\lfloor n/t \rfloor$ and of size $\lfloor n/t \rfloor+1$.

		We run through all $O(k^k)$ possible proper colorings of $X$. For each of these colorings we check whether they can be extended to an equitable coloring of $G$. For a given coloring of $X$, we guess the size of the color class (either $\lfloor n/t \rfloor$ or $\lfloor n/t \rfloor+1$) for each color used in $X$. There are at most $2^k$ possibilities for each coloring of $X$. 
		
		Given a coloring of $X$ with colors from the set $\{1,2,\cdots,k'\}$ where $k'\leq k$, we guess the subsets $Q_C, Q_I \subseteq [k']$ of colors which can be  used to color clique and independent set vertices of $G \setminus X$ respectively in an equitable coloring of $G$ extending the coloring of $X$. We call colors of $Q_C$ as \emph{compulsory} colors of the clique $C$ and colors of $Q_I$ as compulsory colors of the independent set $I$. 
		
		Since $C$ is a clique, each color of $Q_C$ appears exactly once in $C$. However, colors of $Q_I$ may appear more than once in the independent set. As the size of $Q_I$ is at most $k$ and the size of any color class is at most $k+2$,  for each color in $Q_I$ we guess the number times it appears in $I$ in an equitable coloring extending the coloring of $X$.
		
		For each vertex $v$ in the graph $G \setminus X$ we assign a list $L(v)$ of colors from the set $\{1,2, \cdots, k'\}$ based on their neighborhood in $X$ i.e., $ c \in L(v)$ if $v$ is not adjacent to a vertex of color $c$ in $X$. 
		
		Next for each vetex $v$ in the graph $G \setminus X$ we refine the list $L(v)$ in two stages. In the first stage for each vertex $v \in C$, delete all colors from $L(v)$ except the colors present in $Q_C \cap L(v)$ and similarly for  each vertex $v \in I$, delete all colors from $L(v)$ except the colors present in $Q_I \cap L(v)$. 
		
		A color $c$ is \emph{eligible} for a vertex $v$ in the clique if the number of non-neighbors of $v$ in $G\setminus X$ having color $c$ in their lists is greater than or equal to the number of vertices which still need to be colored by the color $c$. For example if a color $c$ appears three times in $X$ and the color class size of $c$ is  $\lfloor n/t \rfloor$ then  $c$ is eligible for a vertex $v\in C$ if the number of non-neighbors of $v$ 
		in $G \setminus X$ having color $c$ in their lists should be at least $\lfloor n/t \rfloor -4$. For each vertex $v \in C$ remove all non-eligible colors from $L(v)$. 
		
		We partition the clique vertices based on their refined list colors, i.e., two vertices $u$ and $v$ belongs to the same set of the partition if $L(u)=L(v)$. Since the size of each list is at most $k$, we can partition vertices of clique into at most $2^k$ subsets. For each subset of the partition, we guess the colors of $Q_C$ which appear in that subset in an equitable coloring.

		We now identify the clique vertices, which can be colored with colors of $Q_C$. 
		In each subset of the partition, we assign the colors of $Q_C$ according to the degree of the vertices inside the threshold graph $G \setminus X$, starting from highest degree to lowest degree. This greedy choice is correct, as threshold graphs satisfy neighborhood containment property. 
		
		For the rest of the clique vertices, we assign new colors from the set $[t] \setminus [k']$. Let $t_1$ be the number of new colors used in the clique, where $t_1 \leq t-k'$. So far, we only know about the sizes of color classes for colors used in $X$ and we don't know the sizes of color classes for the new colors used in the clique. However, using the neighborhood containment property of threshold graphs we can find the color class size for the new colors used in the clique. Let $p$ and $q$ be the number of color classes of sizes 
		$\lfloor n/t \rfloor$ and  $\lfloor n/t \rfloor+1$ respectively after excluding the colors of $X$. We order the vertices of the clique which are colored with new colors according to their neighborhood in $G \setminus X$ from lowest degree to highest degree. Then for the new colors used to color first $q$ vertices in the above ordering we assign their color class size as $\lfloor n/t \rfloor+1$ and 	$\lfloor n/t \rfloor$ for the rest of the colors.

		Now we are only left to color the independent set vertices. Let $t_2=t-t_1-k'$ be the number of colors which are not used in $X$ and $C$. For these $t_2$ colors we can assign the sizes of color classes  based on how many of each size ($\lfloor n/t \rfloor+1$ or 	$\lfloor n/t \rfloor$) still need be covered.  
		To color the independent set vertices we reduce it to a network flow problem as follows.  
		We create a source vertex that is connected  to $t$ vertices representing colors, and the capacity of these
		edges is equal to the number of vertices that still need to be colored by that color. Then we add one vertex for each uncolored vertex in independent set $I$, and edges from these vertices to the sink with capacity one. In the end, we add edges of capacity one from color vertices to the independent set vertices if the color is present in that vertex list of colors. Then we compute a maximum flow and check whether it is equal to the size of the independent set and if this is the case then we 
		immediately obtain an equitable coloring of $G$. 
		
		\paragraph{Running time.}
		For a graph $G$, computing a subset $X$ of size at most $k$ such that $G \setminus X$ is a threshold graph takes $O(4^k (m+n))$ time. We run through all $t \leq r$  number
		of colors for $G$. Then we run through all possible ways of partitioning the vertices of $X$ into color classes of sizes $\lfloor n/t \rfloor$ and $\lfloor n/t \rfloor+1$, which takes time at most $2^{O(k \log k)}$. We guess the colors of $X$ which are used to color some vertices of $G \setminus X$ in time $O(2^k)$. Guessing the compulsory colors in each partition of clique also takes $O(2^k)$ time. We can also guess the $t_1$ and $t_2$ on $O(r)$ time and $p$ and $q$ in $O(t_1)$ time. 
		Finally, we use network flow to decide whether the uncolored vertices of independent set vertices can be equitably colored with respect to the coloring of $X$, and sizes of color classes in this coloring. This takes time $O(mn)$ and the running time of the entire algorithm is $O(2^{k \log k} r^3mn)$.
		
		\end{proof}	
		We showed that \ec{} is $\fpt$ parameterized by distance to threshold graphs. However, the problem is unlikely to admit a polynomial kernel~\cite{bodlaender2010cross}. 
		In the following, we show that when parameterized by $r$ and $k$ the problem admits a polynomial kernel. 

	\begin{lemma}
		\ec{} admits a polynomial kernel parameterized by $r+k$, where $k$ is the distance to threshold graphs and $r$ is the number of colors used. 
	\end{lemma}
	\begin{proof}
		Let $X \subseteq V(G)$ such that $G \setminus X=(C,I)$ is a threshold graph. As $G \setminus X$ has at least one universal vertex, in any \ec{} of $G$ the maximum size a color class is at most $k+2$. Therefore if $n>r(k+2)$ it is a \textsc{No} instance. Otherwise $n \leq r(k+2)$ and the input graph contains at most $r(k+2)$ vertices and at most $O(r^2k^2)$ edges. 
	\end{proof}

\section{List Coloring}
The third variant of graph coloring we study in this paper is \lck{}. Mertzios and Spirakis~\cite{mertzios2013algorithms} showed that 
\textsc {List 3-Coloring} is \NP-complete for graphs of diameter three. The complexity of \textsc {List 3-Coloring} for graphs of diameter two is open. 
In this section, we show that \lck{} is \NP- complete on split graphs, which is a subclass of diameter at most three graphs. Next, we show that the problem is $\fpt$ parameterized by $k$ on split graphs.

\begin{lemma}
	\lck{} coloring is \NP-complete on split graphs. 
\end{lemma}
\begin{proof}
	We give a reduction from the \textsc {Independent Set} problem. Given an instance $(G, k)$ of
	the independent set problem, define a split graph $H=(C,I)$ as follows.
	For every vertex $v \in V(G)$ , we introduce a vertex
	$c_v \in C$. For every edge $uv \in E(G)$ , we introduce a vertex $I_{uv} \in I$ , and connect it with every vertex of $C$  except $c_u$ and $c_v$. We add edges between every pair of vertices in $C$, thereby making $H[C]$ a clique.
	For every vertex $c_v \in C$ assign $L(c_v)=[n]$ and for $I_{uv}\in I$ assign $L(I_{uv})=\{k+1,k+2,\ldots, n\}$.
	
	We can easily see that $H$ is a split graph and it can be constructed in polynomial time. We now show that $G$ contains an independent set of size $k$ if and only if  $H$ is list $k$-colorable.
	
	Suppose that $G$ contains an independent set $X=\{v_{i_1}, v_{i_2}, \ldots, v_{i_k}\}$  of size $k$. Then we can construct a list $k$-colouring of $H$ as follows. 
	\begin{itemize}
		\item Color vertex $c_{v_{i_j}}$ with color $j$, for $1 \leq j \leq k$.
		\item Arbitrarily color the uncolored vertices of the clique with the colors from the set $\{k+1,\ldots, n\}$.
		\item For each $I_{uv} \in I$, at least one of $u$ or $v$ is not in $X$.  This implies that at least one of the colors used for vertices $u$
		or $v$ is from the set $\{k+1, \ldots, n\}$. We color the vertex $I_{uv}$ by the color of $u$ if $v \in X$ and by the color of $v$ otherwise. 
	\end{itemize}
	
	Conversely suppose that $\phi$ is a list $k$-coloring of $H$. Let $X=\{u ~|~ \phi(c_u) \in [k]\}$. We show that $X$ is an independent set of size $k$ in $G$. Clearly the size of $X$ is $k$.  Suppose there exists two vertices  $u, v \in X$ such that $uv \in E(G)$. 
	We know that $\phi(I_{uv})$ is either $\phi(c_u)$ or $\phi (c_v)$. This implies $\phi(I_{uv}) \in [k]$, which is a contradiction to the fact that $L(I_{uv})= \{k+1, \ldots, n\}$. Hence $X$ is an independent set of size $k$ in $G$.	
\end{proof}

	\begin{lemma}
	\lck{} coloring is fixed parameter tractable on split graphs parameterized by $k$.
	\end{lemma}
	\begin{proof}
	Given a split graph $G=(C,I)$ and for each vertex $v \in V(G)$, a list $L(v)$ of $k$ permitted colors. If $|C| > k$ then the given instance is a NO instance as we need at least $|C|$ colors to color the clique. Therefore without loss of generality we assume that $|C|\leq k$. First, we run through the all possible (at most $k^k$) ways of coloring clique vertices with colors from  their lists and check if each such coloring is proper. 
	Then we try to extend each proper coloring of clique $C$ to the rest of $G$ as follows.
	For each $v \in I$, color it with any color from $L(v)$ which is not used to color any vertex of $N(v)$. Altogether this gives an $O(k^k(m+n))$ FPT algorithm.	
	\end{proof}

\section{Conclusion}

In this paper, we study the parameterized complexity of several graph coloring problems. We showed that (a) \pe{} and \ec{} are $\fpt$ parameterized by distance to threshold graphs and (b) \lck{} is $\fpt$ parameterized by $k$ on split graphs. 

The following are some interesting open problems.
\begin{enumerate}
	\item What is the complexity of \lck{} for (a) threshold graphs (b) complete-split graphs (c) diamter two graphs.
	\item What is the parameterized complexity of \textsc{Number Coloring}~\cite{doucha2012cluster} (generalization of \ec{}) parameterized by distance to threshold graphs?
	\item It is known that \gc{} admits polynomial kernel parameterized by distance to clique~\cite{das2017structural}. Does \pe{} and \ec{} admit polynomial kernel parameterized by distance to clique?
\end{enumerate}
\bibliographystyle{splncs04}
\bibliography{equitable}

\end{document}